\def\year{2022}\relax
\documentclass[letterpaper]{article} 
\usepackage{aaai22}  
\usepackage{times}  
\usepackage{helvet}  
\usepackage{courier}  
\usepackage[hyphens]{url}  
\usepackage{graphicx} 
\usepackage{natbib}  
\usepackage{caption} 
\DeclareCaptionStyle{ruled}{labelfont=normalfont,labelsep=colon,strut=off} 
\usepackage{algorithm}
\usepackage{algorithmic}

\usepackage{newfloat}
\usepackage{listings}
\floatstyle{ruled}
\newfloat{listing}{tb}{lst}{}
\floatname{listing}{Listing}
\usepackage{microtype}
\usepackage{subfigure}
\usepackage{amsmath}

\DeclareMathOperator*{\argmin}{argmin}
\usepackage{booktabs}
\usepackage{bm}
\usepackage{amsthm}
\usepackage{mathtools}
\usepackage{amsbsy}
\usepackage{amssymb}
\usepackage{cleveref}
\def \bbP {\mathbb{P}}

\def \bbQ {\mathbb{Q}}
\def \bx {\bm x}

\def \cX {\mathcal{X}}
\def \cY {\mathcal{Y}}
\def \cM {\mathcal{M}}
\def \cD {\mathcal{D}}
\def \cF {\mathcal{F}}
\newtheorem{assumption}{Assumption}
\newtheorem{remark}{Remark}

\title{Continual Learning for CTR Prediction: A  Hybrid Approach}
\author{
Ke Hu$^*$, Yi Qi$^*$, Jianqiang Huang, Jia Cheng, Jun Lei
}
\affiliations{
    Meituan \\
    Beijing, China \\
    \{huke05,qiyi02,huangjianqiang,jia.cheng.sh,leijun\}@meituan.com

}

\begin{document}

\maketitle

\begin{abstract}
Click-through rate(CTR) prediction is a core task in cost-per-click(CPC) advertising systems and has been studied extensively by machine learning practitioners. While many existing methods have been successfully deployed in practice, most of them are built upon i.i.d.(independent and identically distributed) assumption, ignoring that the click data used for training and inference is collected through time and is intrinsically non-stationary and drifting. This mismatch will inevitably lead to sub-optimal performance. To address this problem, we formulate CTR prediction as a continual learning task and propose COLF, a  hybrid \textbf{CO}ntinual \textbf{L}earning \textbf{F}ramework for CTR prediction, which has a memory-based modular architecture  that is designed to adapt, learn and give predictions continuously when faced with non-stationary drifting click data streams. Married with a memory population method that explicitly controls the discrepancy between memory and target data, COLF is able to gain positive knowledge from its historical experience and makes improved CTR predictions. Empirical evaluations on click log collected from a major shopping app in China demonstrate our method's superiority over existing methods. Additionally, we have deployed our method online and observed significant CTR and revenue improvement, which further demonstrates our method's efficacy. 
\end{abstract}

\section{Introduction}
\begin{figure}[ht]
\begin{center}
    \includegraphics[width=0.9\linewidth]{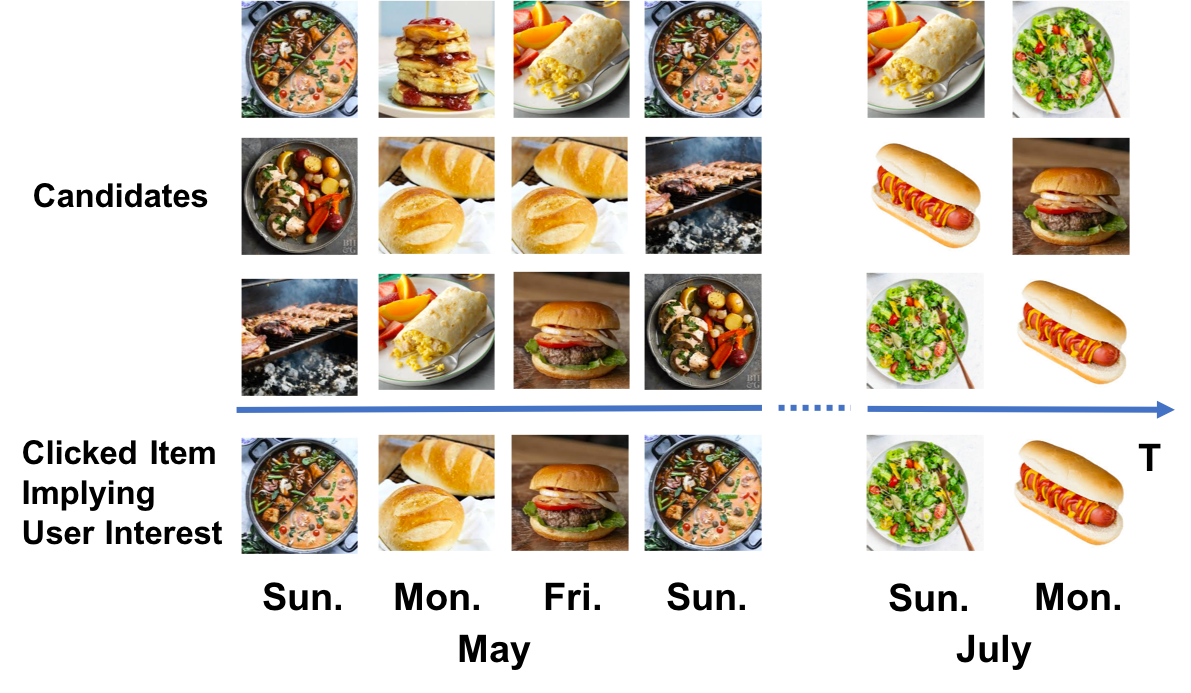}
    \caption{A visual example of the non-stationary  and drifting pattern of click log. On the one hand, candidates and clicked items that imply user's interest are quite different between two remote times, e.g., between May and July. On the other hand, candidates and clicked items are alike in adjacent times, such as the repeated click on hotpot on Sunday in May and the recurring candidate items in May. }
    \label{fig:inspiringFig}
    \end{center}
\end{figure}
Cost-per-click(CPC) advertising systems have achieved huge commercial success over the past decade and are ubiquitous nowadays. In CPC advertising systems, advertisements are ranked by eCPM, i.e., the product of the bid price and the predicted click-through rate(CTR). To guarantee reliable service quality and maximize revenue, these ranking systems rely heavily on accurate prediction of advertisements' click-through rate. Recent years have seen fruitful progress made by developing sophisticated deep models to improve CTR prediction's quality, such as  neural model combining factorization machine \cite{guo2017deepfm} and attention-based neural model \cite{feng2019deep}. These models not only obtained improvements  in offline experiments but also raised CTR and revenue in online A/B test.

In spite of the rapid development of CTR model's architectures, the foundation of those models has rarely been investigated. The click log, which is used as training data for these machine learning methods, is collected sequentially through time and is intrinsically time-varying. It is common that changes of available  advertisements and changes of user interest make users' click behavior vary over time. As Figure \ref{fig:inspiringFig} shows, on a Monday in May, bread advertisement was clicked while on a Monday in July, hot dog was clicked when bread advertisement is unavailable(due to leave of advertiser, for example). Note that these changes usually happen slowly and there are also recurring click patterns within a short time. We say that the distribution of click log is \textbf{\textit{non-stationary and drifting}} in that available candidates and users' interest stay alike within a short time but will eventually become quite different after a long time. However, most existing methods on CTR prediction ignore the non-stationary and drifting characteristics of click log and implicitly assume a stationary distribution of the training data. This modeling pitfall inevitably leads to sub-optimal performance, especially when neural models are under consideration, since neural models are prone to catastrophic forgetting and negative knowledge transfer when faced with non-stationary data \cite{hadsell2020embracing}. 

Recently, continual learning has attracted much attention in building adaptive systems that are able to gain, retain and transfer knowledge when faced with non-stationary data streams. The main goal of continual learning is to mitigate catastrophic forgetting as well as to foster positive knowledge transfer in the continuous learning process. Despite that the non-stationary and drifting characteristics of click data calls for special treatment to catastrophic forgetting and knowledge transfer, the continual learning approach for CTR prediction remains unexplored. It is a natural question to ask, whether CTR prediction algorithms could benefit from continual learning techniques and obtain more precise CTR predictions in this non-stationary world?

In this work, we explore the continual learning approach for CTR prediction and answer the question above positively. We regard CTR prediction faced with non-stationary and drifting click log as a sequence of learning tasks identified by time, e.g. by date. We propose COLF, a \textbf{CO}ntinual \textbf{L}earning \textbf{F}ramework for CTR prediction that is able to adapt, learn and give predictions continuously through a hybrid approach, which consists of a modular architecture to avoid negative impact between tasks and a memory replay module to mitigate catastrophic forgetting  and foster positive knowledge transfer. As for memory replay to achieve so, we introduce a special memory population method that explicitly controls the discrepancy between memory and target data. We test COLF on a large-scale real world dataset and obtain significant improvement over existing methods. We have also deployed COLF online and rigorous A/B test showed substantial CTR and revenue boost, which further demonstrates our method's efficacy.

\subsection{Document Structure}
In Section \ref{section:relatedWork} we review existing works relevant to ours. In Section \ref{l:problem-setup} we present our formulation and assumptions of continual learning for CTR prediction. In Section \ref{section:method} we present the proposed hybrid method COLF. In Section \ref{section:exp} we describe our experimental work and discuss the result. Finally, we present the concluding remark in Section \ref{section:conclusion}.

\section{Related Work}
\label{section:relatedWork}
Our work lies in the intermediate area of two lines of research, i.e., CTR prediction and continual learning. 
\subsection{CTR Prediction}

CTR(click-through rate) prediction is a core task in cost-per-click (CPC) advertising systems, where items are ranked by the product of  bid price and predicted CTR. The precision of CTR prediction model is crucial to systems' success. Substantial efforts have been made on the design of model architectures in research works on CTR prediction.

LR(logistic regression) and GBDT are two classical  models for CTR prediction and had been widely adopted in industry, such as Google Ads \cite{mcmahan2013ad} and Facebook Ads \cite{he2014practical}. Recent years have seen rapid development in neural model's application in CTR prediction, for example, the Wide\&Deep model \cite{cheng2016wide}, which could be seen as a combination of deep neural model and LR model.
Following Wide\&Deep, neural models become the \textit{de facto} approaches to CTR prediction. Some representative models are DeepFM \cite{guo2017deepfm}, DIN \cite{zhou2018deep} and DSIN \cite{feng2019deep}. 

All these models assume implicitly  that the training data's distribution is i.i.d. and have sub-optimal performance in  real world applications. There are very few work on CTR prediction addressing the non-stationary and drifting data problem. A recent work studied session-based recommendation task in continual learning setting by utilizing memory-based method to mitigate catastrophic forgetting \cite{mi2020ader}. However, it gave no formal treatment to the non-stationary and drifting pattern of real world data  and relied on  heuristics to populate memory, with no emphasis on positive knowledge transfer, which is opposite to our method. 

\subsection{Continual Learning}
Most powerful modern machine learning algorithms perform well only when the presented data is stationary. However, the world is intrinsically non-stationary. Continual learning is an increasingly relevant research area that try to find ways for machine learning models to learn sequentially from  non-stationary data \cite{lopez2017gradient,hadsell2020embracing}.

There are three main paradigms of continual learning, which are regularization-based methods, methods using modular architecture and memory-based methods. Regularization-based  methods force the gradient on new task to stay aligned with gradients from previous learned tasks \cite{lopez2017gradient,chaudhry2018efficient} or the newly learned parameters to fluctuate minimally from the old model so to avoid catastrophic forgetting \cite{li2017learning,kirkpatrick2017overcoming,zenke2017continual}.
The main weakness of regularization-based methods is the strict limitation of model capacity which may results in poor adaptation to new data. 

Methods using modular architecture  grow the base neural network as required when faced with newly arrived data in order to avoid negative impact on learnt knowledge. Knowledge transfer from the past to the future is realised by sharing some bottom layers or low-level feature representations \cite{serra2018overcoming,mallya2018packnet,li2019learn}. The potential problem of these methods is that  the ability of knowledge transfer is limited and the model size may grow too large. 

Memory-based methods construct a memory set to store the learnt knowledge for future use \cite{robins1995catastrophic,riemer2018learning,sprechmann2018memory,isele2018selective,rolnick2019experience}. 
When the memory is filled with sampled historical data, the content of the memory is also called exemplars, and the technique to take advantage of the memory is called replay, which means training on the memory as if recalling the historical experience. Many works focus on how to select the most representative and effective exemplars to populate the memory \cite{aljundi2019online,guo2020improved}.
Memory-based approaches are shown to be more reliable than regularization-based methods \cite{knoblauch2020optimal}. Our method takes a hybrid approach that marries modular architecture with memory replay, enjoying the advantages of these two paradigms while avoiding the weakness of both.
\section{Problem Setup}
\label{l:problem-setup}
We consider CTR prediction  in the continual learning setting. Throughout, we denote $\cX$ as the input space. We denote $\bx = (u, v, c) \in \cX$ as the usual input to a CTR prediction model, where $u$ is the user, $v$ is the target item(e.g., advertisement in advertising systems) and $c$ is the context information. Denote the binary random variable of observed click by $y\in Y$, where $y = 0$ indicates no click and $y=1$ indicates a click. Define $\cF$ as the hypothesis space which is always realised by neural models of a classical embedding-dense architecture like Wide\&Deep \cite{cheng2016wide}(See the right part of Figure \ref{fig:modularArchitecture}).

A standard CTR prediction task is defined by the learning problem of conditional click probability estimator
\begin{equation}
\bbP(y=1|\bx,(\mathcal{D},\cY)) = \hat{f}(\bx)(\bx\sim\bbQ)
\end{equation}
where $(\cD,\cY)=\{(\bx_i,y_i)\}_{i=1}^N$ is the observed click dataset that is sampled from $(\cX,Y)$ according to some distribution $\bbQ$. We assume that there is a ground-truth click probability function and denote it by $\bar{f}$.
Under i.i.d. assumption, the estimator is usually given by the minimization of accumulated cross-entropy loss, i.e.,
\begin{equation}
\label{originalCTRTask}
        \hat{f}(\bx) = \argmin_{f\in\cF}  \sum_{(\bx_i,y_i)\in(\cD,\cY)}l(f,\bx_i,y_i)
\end{equation}
where $l(f,\bx_i,y_i)=y_i\log f(\bx_i)  + (1-y_i)\log (1 - f(\bx_i))$ is the cross-entropy loss of a single sample. 

In real world, both the input dataset $\cD$ and the mapping function $\bar{f}$ from input $\cD$ to output $\cY$ are time-varying. Denote the dataset sequence by $(\cD_1,\cY_1),(\cD_2,\cY_2),...(\cD_t,\cY_t)(t\in N,t\rightarrow \infty)$, the corresponding sampling distributions by $\bbQ_1,\bbQ_2,...\bbQ_t$ and the corresponding ground-truth mapping functions by $\bar{f}_1,\bar{f}_2,...\bar{f}_t$. The CTR prediction task in the continual learning setting is the continuous learning tasks of the click probability function $\hat{f}_t$ over time $t$. Given a specific time $t$, the learning task is defined by 
\begin{equation}
    \bbP(y=1|\bx,\{(\cD_i,\cY_i)\}_{i=1}^{t-1}) = \hat{f}_t(\bx)(\bx\sim\bbQ_t)
\end{equation}
where the condition on $\{(\cD_i,\cY_i)\}_{i=1}^{t-1}$ indicates the dependency on all historical data observed so far and the target on $\bbQ_t$ indicates the gap between training data's distribution and test data's. It's easy to see that the standard CTR prediction task is a special case of the the continual CTR prediction task, where all dataset $\mathcal{D}_i$ are sampled from the same distribution, i.e., $\bbQ=\bbQ_1=\bbQ_2=...=\bbQ_{t}$ and all conditional click probability functions $\bar{f}_i$ are the same $f$. However, when $\bbQ_i$ and $\bar{f}_i$ are time-varying, reusing Equation \ref{originalCTRTask} at every time step $t$ as the estimator results in sub-optimal performance since usually 
\begin{equation}
\begin{aligned}
    & \argmin_{f\in\cF}\sum_{j=1}^{t-1}\sum_{(\bx_i,y_i)\in(\cD_j,\cY_j)}l(f,\bx_i,y_i) \\
    & \neq \argmin_{f\in\cF}\sum_{(\bx_i,y_i)\in(\cD_t,\cY_t)}l(f,\bx_i,y_i).
\end{aligned}
\end{equation}
To tackle this problem, we make two reasonable assumptions on how $(\bbQ_i,\bar{f}_i)$, i.e., the distribution of $(\cD_i,\cY_i)$ may drift over time.
\begin{assumption}
\label{assumption:dataDistribution}
The distributions $\{\bbQ_i\}_{i=1}^t$ are similar in adjacent periods, but quite different when there is a  long time gap. To be more specific, we assume
    $KL(\bbQ_i, \bbQ_{i+1})< \epsilon_q$ and $KL(\bbQ_i,\bbQ_{k})> \epsilon_q,\forall k>i+m_q$, where $KL$ is the K-L divergence between two distributions and $\epsilon_q,m_q$ are problem-related thresholds.
\end{assumption}
\begin{assumption}
\label{assumption:clickFunction}
The click probability functions $\{\bar{f}_i\}_{i=1}^t$ are similar in adjacent periods, but quite different when there is a  long time gap. To be more specific, we assume
$C(\bar{f}_{i+1},\bbQ_{i+1}) - C(\bar{f}_i,\bbQ_{i+1})< \epsilon_f$ and $C(\bar{f}_{i},\bbQ_{i}) - C(\bar{f}_k,\bbQ_{i})   > \epsilon_f,\forall k>i+m_f$, where $C(f,\bbQ)$ is a problem-related criteria to evaluate $f$'s performance on dataset sampled from $\bbQ$, and $\epsilon_f,m_f$ are problem-related thresholds.
\end{assumption}

Figure \ref{fig:inspiringFig} is an illustrative figure on the two assumptions. We also validate these assumptions on real world data collected from a major shopping app from China(See Section \ref{exp:dataset} for details). Note that our assumptions do not assume a  mono-directional change of $(\bbQ_i,\bar{f}_i)$. Rather, it only restrict the speed of change  within a short time and assumes a significant change  in long terms, which leaves space for continual learning methods to mitigate forgetting of useful knowledge in recent times thus fostering positive knowledge transfer. We introduce our continual learning method in next section.

\begin{remark}
It is most common in practice to divide click log  into partitions by date. Note that for some applications like news recommender systems, partitioning by hour may be a more suitable criteria due to its sensitivity to time.
\end{remark}

\begin{figure*}[ht]
\begin{center}
    \includegraphics[width=\linewidth]{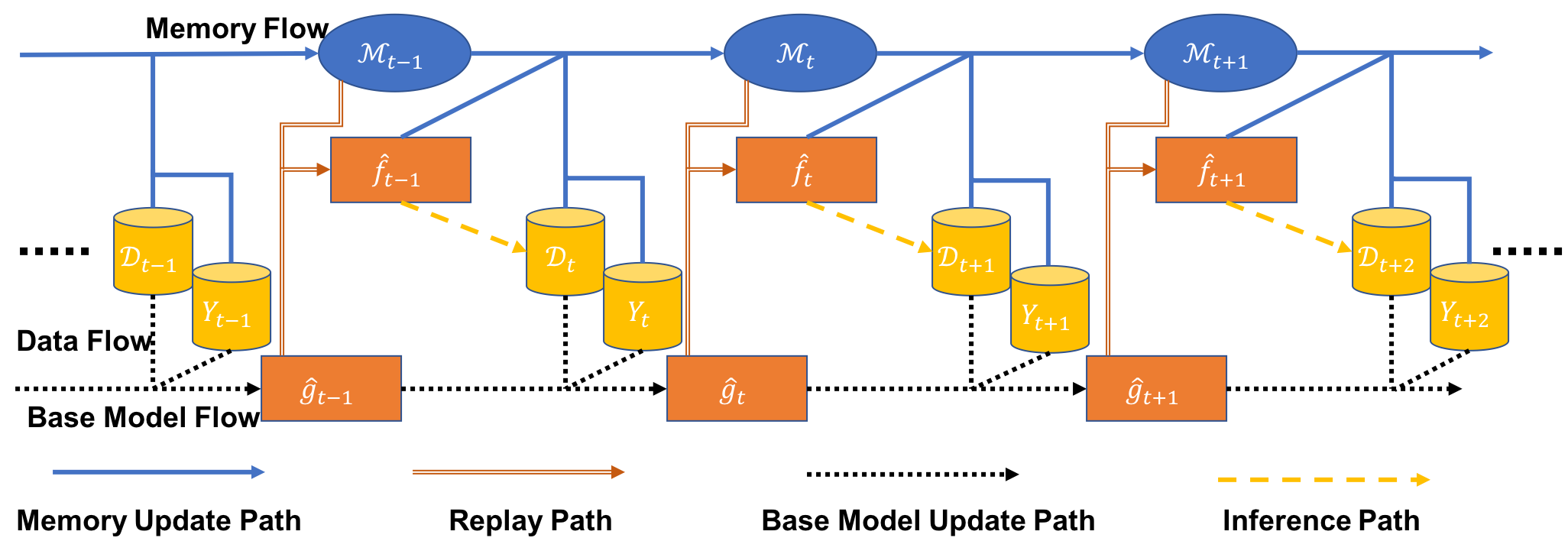}
    \caption{The workflow of COLF. Three main information flows over time are annotated, and the arrow lines indicate dependencies between modules. Note that the actual inference on $\cD_{t+1}$ is done by $\hat{f}_t$, which relies on the base model $\hat{g}_t$ and the memory $\cM_t$.}
    \label{fig:algorithmFlowFig}
    \end{center}
\end{figure*}
\section{Method}
\label{section:method}
In continual learning for CTR prediction, the learner has to estimate $\hat{f}_t$ only with labeled data before time $t$, i.e., $\{(\mathcal{D}_i,\cY_i)\}_{i=1}^{t-1}$. To accomplish this goal, the main intuition is to construct a training dataset approximating to $(\mathcal{D}_{t},\cY_t)$ and learning from an approximated function of $\bar{f}_t$. Based on this intuition, we propose COLF,  a hybrid  \textbf{CO}ntinual \textbf{L}earning \text{F}ramework for CTR prediction. In this section, we first introduce how COLF does learning and inference continuously over time by utilizing a base model flow in company with a memory flow under a modular architecture. Then we present our memory population method that controls distribution discrepancy between memory and target data. We give a complete description of our method in the end of this section.
\subsection{The Workflow of COLF}
The workflow of COLF is illustrated in Figure \ref{fig:algorithmFlowFig}. COLF contains two flows $\hat{g}_t(t\in N_+)$ and $\cM_t(t\in N_+)$ in company with the external data flow $\{D_i,Y_i\}_{i=1}^\infty$ and a growing modular architecture $\{\hat{f}_t\}(t\in N_+)$ used for actual inference. The first flow is the base model flow $\hat{g}_t$, which takes the role of consuming newly arrived data at each time $t$ to adapt quickly to the distribution $\bbQ_t$. When the data flow  $(\cD_t, \cY_t)\sim(\bbQ_t,\bar{f}_t)$ is fully observed, COLF updates $\hat{g}_{t-1}$ by 
\begin{equation}
    \label{equation:updateOfg}
    \hat{g}_t = \argmin_{g\in\cF} \sum_{(\bx,y)\in(\cD_t,\cY_t)} l(g,\bx,y)
\end{equation}.
This update is depicted by base model update path in Figure \ref{fig:algorithmFlowFig}. The second flow is the memory flow $\cM_t$, which is responsible for storing historical exemplars that are  similar to distribution $\bbQ_{t+1}$, whose update mechanism is to be explained in detail in the next section. Note that memory $\cM_t$ is dependent on $(\cM_{t-1},\hat{f}_{t-1},(\cD_t,\cY_t))$, where $\hat{f}_{t-1}$ is the mapping function that is actually used for inference to give CTR predictions. The memory update is depicted by memory udpate path in Figure \ref{fig:algorithmFlowFig}. 

\begin{figure}[ht]
\begin{center}
    \includegraphics[width=\linewidth]{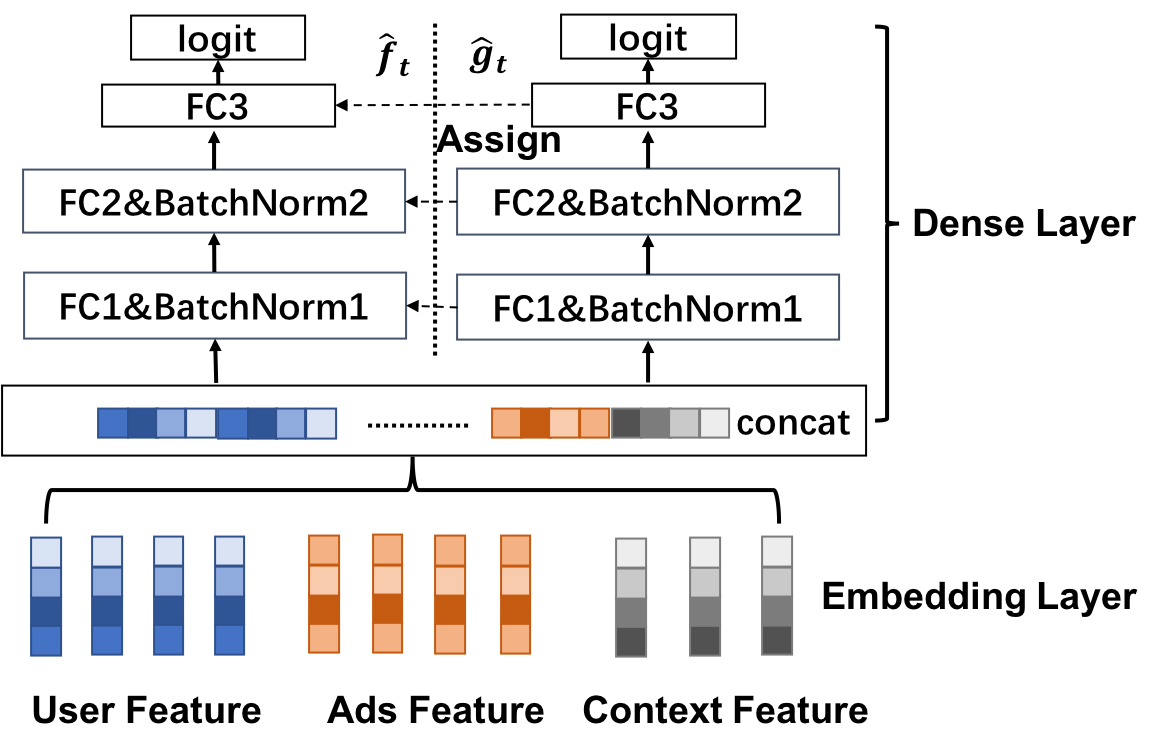}
    \caption{Architecture of base model $\hat{g}_t$(the right part) and the training of modular architecture of $\hat{f}_t$(the left part).}
    \label{fig:modularArchitecture}
    \end{center}
\end{figure}

Now we introduce $\hat{f}_t$. Since $\cM_t$ is expected to be similar to the upcoming target data $(\cD_{t+1},\cY_{t+1})$ and $\hat{g}_t$ is an estimator of $\bar{f}_t$ which is similar to $\bar{f}_{t+1}$ according to Assumption \ref{assumption:clickFunction}, it is straightforward to train $\hat{f}_t$ by 
\begin{equation}
\label{equation:f}
    \hat{f}_t = \argmin_{f\in\cF} \sum_{(\bx,y)\in\cM_t}  l(f,\bx,y) + L(\hat{g}_t,f,\bx,y).
\end{equation}
On the one hand, we decoupled $\hat{f}_t$ from $\hat{g}_t$ to avoid negative impact on subsequent tasks. On the other hand, to facilitate knowledge transfer, we use a modular architecture based on $\hat{g}_t$ for $\hat{f}_t$. As is illustrated in Figure \ref{fig:modularArchitecture}, $\hat{f}_t$ and $\hat{g}_t$ share the same bottom embedding layer while $\hat{f}_t$ grows its own dense layer. We initialize weights of  $\hat{f}_t$'s dense layer by $\hat{g}_t$'s to improve convergence rate.  
Once  $\cD_{t+1}$ is observed, $\hat{f}_t$ will give CTR predictions on $\cD_{t+1}$ for downstream tasks (e.g. eCPM calculation). When  $\cY_{t+1}$ of $\cD_{t+1}$ is fully observed, the cycle of update, training and inference above  begins again. The working path of $\hat{f}_{t}$ is depicted by replay path and  inference path in Figure \ref{fig:algorithmFlowFig}.
\subsection{Memory Population Method}
Partition $\cM_{t-1}$ into $ (\cM_{t-1}^{t-1-k_t},\cM_{t-1}^{t-k_t}, ...\cM_{t-1}^{t-1})$ according to the date of data collection, where $\cM_{t-1}^{t-j}$ is the subset of $\cM_{t-1}$ that contains all exemplars collected at time $t-j$ and $t-1-k_t$ is the oldest time of exemplars in $\cM_{t-1}$. We introduce our memory population method to update $\cM_{t-1}$ to $\cM_t$ that explicitly controls distribution discrepancy between $\cM_t$ and  $(\cD_{t+1},\cY_{t+1})$ now.
There are three steps to achieve the above goal, which are discarding old memory, refreshing relevant memory and appending  new memory.

\textbf{Discarding old memory.} According to Assumption \ref{assumption:clickFunction}, $\bar{f}_t$ and $\bar{f}_{t-k}$ are quite different when $k$ is large but are relatively close when $k$ is small. We  can identify the older part in $\cM_{t-1}$ according to the corresponding $\hat{f}$'s performance on the latest data $(\cD_t,\cY_t)$, which is used as an approximation to $(\bbQ_{t+1},\bar{f}_{t+1})$ which we  know nothing about at $t$. To be more specific, a memory partition  $\cM_t^{t-i}$ is said to be old if 
\begin{equation}
    \textbf{1}_{t-k}\{C(\hat{f}_{t-1},(\cD_t,\cY_t))- C(\hat{f}_{t-i},(\cD_t,\cY_t)) > \epsilon\}
    \label{oldmemory}
\end{equation}
equals 1, where $\textbf{1}$ is the index function, $C(f,(\cD,\cY))$ is any function that  evaluates $f$'s performance on $(\cD,\cY)$ and $\epsilon$ is the  threshold of old memory surviving. A typical choice of $C$ could be the AUC score and $\epsilon$ is problem-related. See Section \ref{section:exp} for more discussion. The old memory of $\cM_t$ is given by
\begin{equation}
    \cM_{t-1}^{old} = \cup_{1\leq k\leq  k_t+1, \textbf{1}_{t-k}=1} \cM_{t-1}^{t-k} 
\end{equation}
which should be discarded to avoid negative transfer since their corresponding $\hat{f}$ are far away from the wanted $\bar{f}_{t+1}$.

\textbf{Refreshing relevant memory.} We denote $\bar{\cM}_{t-1}=\cM_{t-1} - \cM_{t-1}^{old}$. We further identify the relevant samples $(\bx,y)\in\bar{\cM}_{t-1}$ by its likelihood to be in $\bbQ_{t+1}$. Again, we use $\cD_t$ as an approximation to $\bbQ_{t+1}$. The likelihood function is a maximum likelihood estimator trained on $\cD_t$ and is denoted by $\hat{p}_t$. A typical choice of $\hat{p}_t$ could be the function of item frequency in $\cD_t$, since items are given by the information system and have a natural ID to count on. The irrelevant memory set is defined by 
\begin{equation}
    \cM_{t-1}^{irrelevant} = \{(\bx,y)|(\bx,y)\in\bar{\cM}_{t-1},\hat{p}_t(\bx)<\epsilon_p\}
    \label{equation:irrelevant}
\end{equation}
where $\epsilon_p$ is the surviving threshold of irrelevant memory and the relevant memory is given by $\cM_{t-1}^{relevant} = \bar{\cM_t} -\cM_{t-1}^{irrelevant} $. The choice of $\epsilon_p$ is problem dependant. See Section \ref{section:exp:continuallearning} for more discussion. $\cM_{t-1}^{relevant}$ should be kept to avoid catastrophic forgetting of useful knowledge as well as to foster positive knowledge transfer.

\textbf{Appending new memory.} Denote $\cM_t^{New} = (\cD_t,\cY_t)$. In addition to the refreshed memory, we append new exemplars from $(\cD_t,\cY_t)$ to the memory set. Since we assume both the data distributions and the click probability functions are very similar in adjacent periods, we append the whole $\cM_t^{new}$ to the memory $\cM_t$.

As a result, we have 
\begin{equation}
\begin{aligned}
        \cM_{t} &=  \cM_t^{new}\cup\cM_t^{relevant} \\
        &=\cM_{t-1} - \cM_{t-1}^{old} - \cM_{t-1}^{irrelevant} + \cM_t^{new}.
        \label{equation:updateMemory}
\end{aligned}
\end{equation}

\subsection{Discussion of COLF}
Combining the modular architecture and the memory population method introduced in previous sections, we have COLF as is described in Algorithm \ref{algorithm}.
\begin{algorithm}[ht]
  \caption{COLF}\label{algorithm}
  \begin{algorithmic}[1]
  \item Initialize the base model $\hat{g}_0,\hat{f}_0\in \mathcal{F}$, the initial memory set $\mathcal{M}_0 = \emptyset$.
  \FOR{$t$=1,2,3,...$T$,...}
  \STATE Observe $\cD_t$.
  \STATE Give predictions on $\cD_t$ using $\hat{f}_{t-1}$.
  \STATE Observe $\cY_t$, the label of $\cD_t$ and evaluate the performance of $\hat{f}_{t-1}$.
  \STATE Update the base model and get $\hat{g}_t$ based on $\hat{g}_{t-1},(\cD_t,\cY_t)$ by Equation \ref{equation:updateOfg}.
  \STATE Update the memory module and get $\cM_t$ based on $\cM_{t-1},(\cD_t,\cY_t)$ by Equation \ref{equation:updateMemory}.
  \STATE Update the evaluation function and get $\hat{f}_t$ based on  $\cM_t$ by Equation \ref{equation:f} with a modular architecture sharing with $\hat{g}_t$ as is illustrated in Figure \ref{fig:modularArchitecture}.
  \ENDFOR
  \end{algorithmic}
\end{algorithm}

It has been established that continual learning with arbitrary data stream  is almost always NP-hard \cite{knoblauch2020optimal}. Despite that we have restrictions on the data distribution, the optimal continual learning under our setting is still NP-hard, given that  $\bar{f}_i$, which is approximated by neural network, has quite complex geometric shape, which is also discussed in the paper above. Though  being an approximation method, COLF is expected to outperform existing methods due to its tailored treatment to the non-stationary and drifting characteristics of click data. We present our experimental results in detail in the next section.
\section{Experiment}
\label{section:exp}
\subsection{Experimental Dataset}
\label{exp:dataset}
The experimental dataset is collected from a major shopping mobile app in China during a one-year period. Every record in the data has a timestamp identifying its actual event time. We partition the dataset into different parts by date. The statistics of the sampled dataset is summarized in Table \ref{table:dataCharacteristics}. 

\begin{figure}[ht]
\begin{center}
    \includegraphics[width=\linewidth]{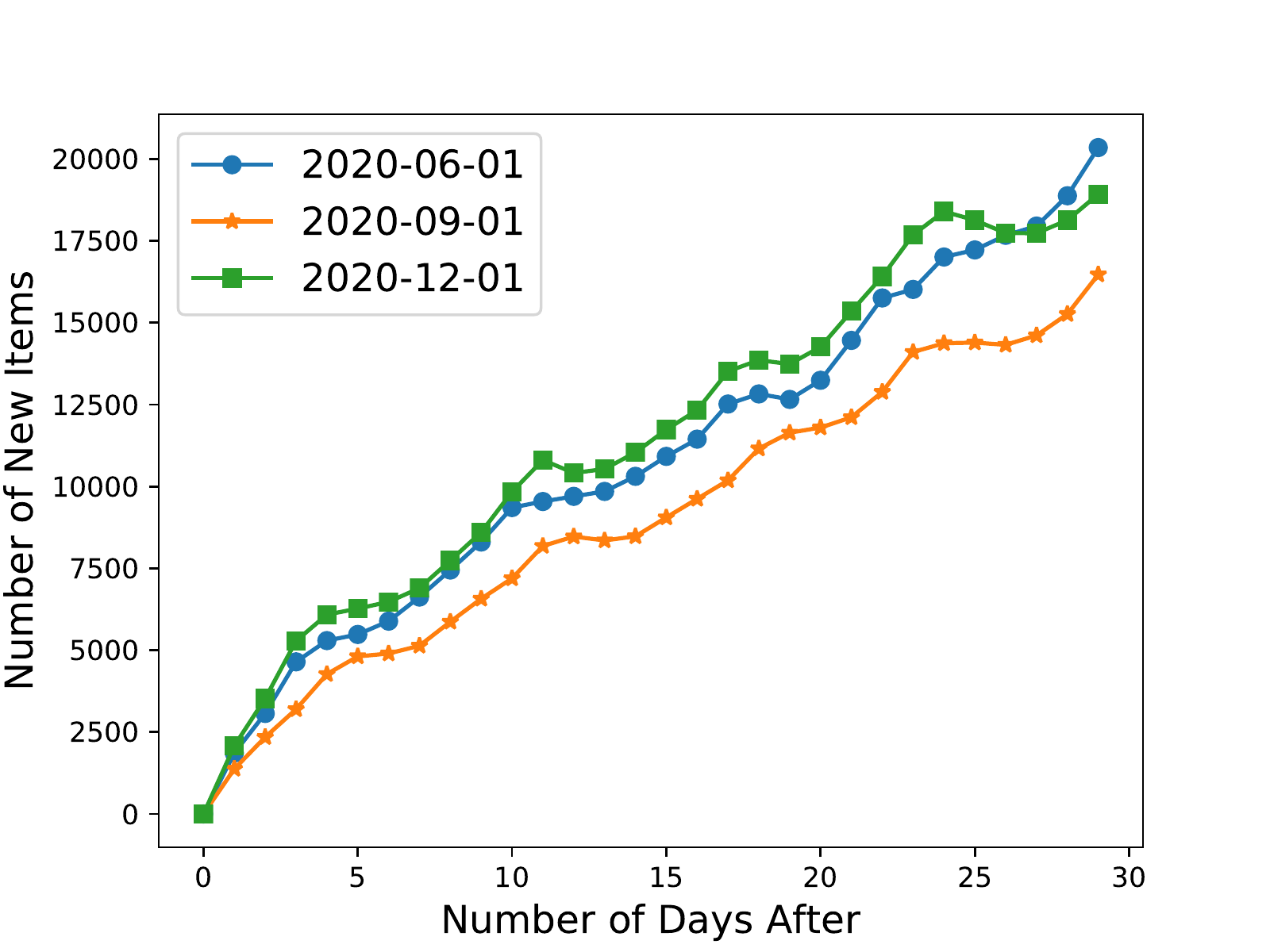}
    \caption{Drift of item set. The number of new items is small within a short time but tends to increase over time. The number reaches a very high level after a long time, regardless of which date is chosen as base.}
    \label{fig:data_distribution}
\end{center}
\end{figure}

\begin{figure}[ht]
\begin{center}
    \includegraphics[width=\linewidth]{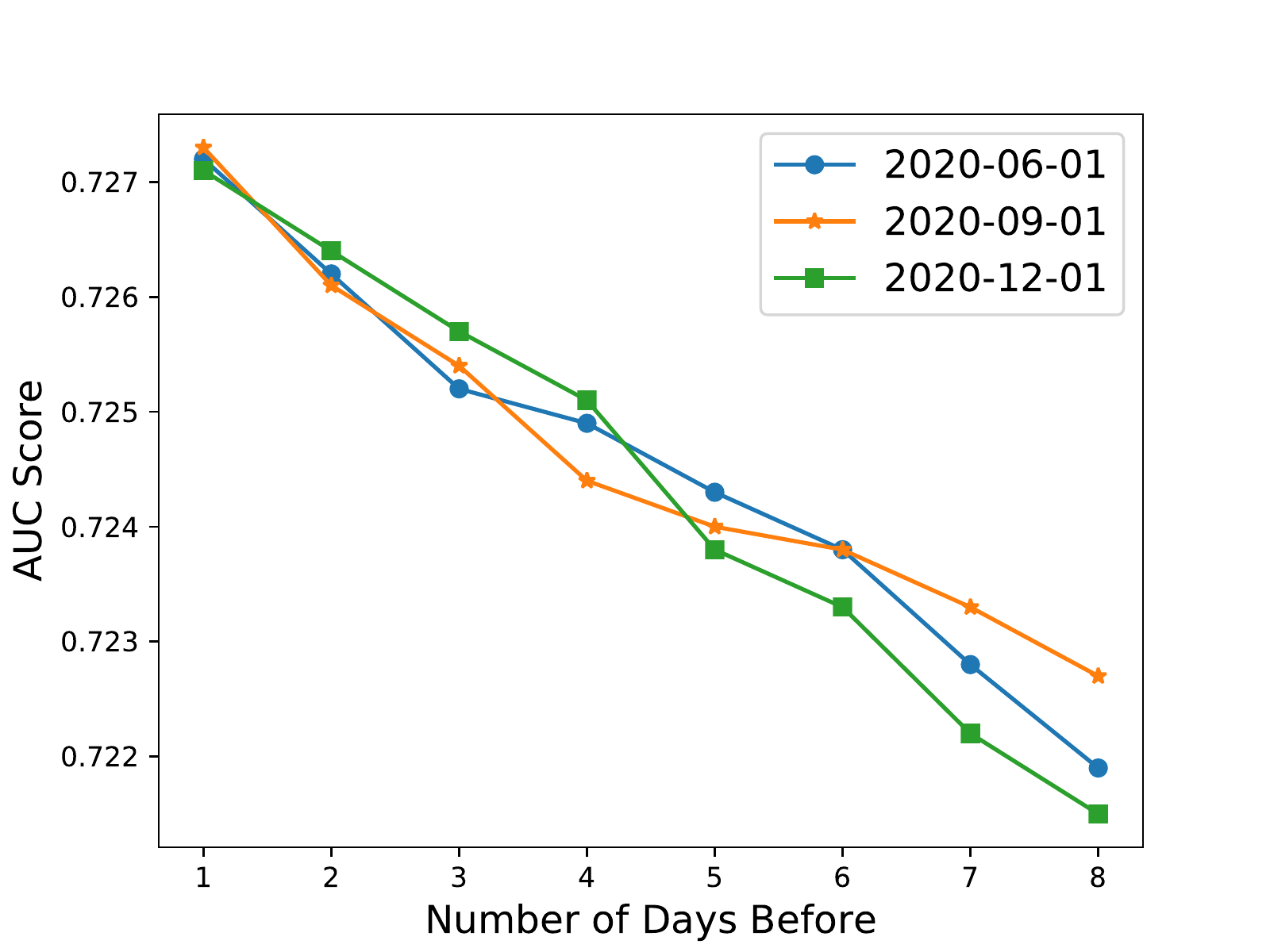}
    \caption{Drift of click probability function. AUC score drops slightly within a short time but will deteriorate  sharply as the time gap between last day of training data and the date of test data becomes large regardless of which date is chosen as base.}
    \label{fig:auclist}
    \end{center}
\end{figure}

Figure \ref{fig:data_distribution} shows that in adjacent periods, number of new items is small while after a long period the number become much larger, which is in accordance to Assumption \ref{assumption:dataDistribution}.  Figure \ref{fig:auclist} shows that Assumption \ref{assumption:clickFunction} holds in real world data. In adjacent periods, the AUC score is quite close whatever training date is chosen, but with larger time gap, the performance becomes poorer and eventually drops 0.5\%.
\begin{table}[t]
\caption{Statistics of the dataset used for offline experiment.}
\label{table:dataCharacteristics}
\begin{center}
\begin{small}
\begin{tabular}{lr}
\toprule
 & Number \\
\midrule
Total Data Size&  $\sim 3$ billion \\
Average Size per Day  & $\sim 10$ million \\
Feature Num & $\sim$ 150 \\
Total Items & $\sim 1$ million \\
Average Items per Day & $\sim 200$ thousand \\
\bottomrule
\end{tabular}
\end{small}
\end{center}
\end{table}
\subsection{Continual Learning Algorithms for Comparison}
\label{section:exp:continuallearning}
We introduce the continual learning algorithms used for comparison in this section. Note that the memory size of all methods is restricted by 70 million (the average size per week) for fair comparison.

\textbf{CBRS.} Class-balancing reservoir sampling(CBRS) is a method proposed to learn continually with temporally correlated and severely imbalanced data \cite{chrysakis2020online}, which is exactly the case of click log. It fills and updates a fixed-size memory for replay such that  all classes are equally distributed to the maximum extend. Different from it, COLF has a different memory population method that explicitly controls the discrepancy between memory and target data.

\textbf{ADER} ADER is a continual learning algorithm targeting at session-based recommendation task \cite{mi2020ader}.  It updates memory  according to items' historical frequency only, thus may suffer from insufficient positive transfer. We include it here to show that our memory population strategy's advantage. 

\textbf{COLF} COLF is the method we propose in this paper that use a hybrid continual learning approach based on modular architecture and memory replay for CTR prediction. We choose  $\epsilon$ in Equation \ref{oldmemory} to be 0.003 since a 0.003 difference in AUC score will result in significant different online performance according to past experience. We choose $\hat{p}_t$  in Equation \ref{equation:irrelevant} to be the item frequency function and $\epsilon_p$ to be 1e-6, guaranteeing a minimal occurrence of 100 times in the memory (10 million is the average size of our data per day. see Table \ref{table:dataCharacteristics}), since a minimal number of 100 occurrences is sufficient to learn a good item embedding according to our past experience. We show that COLF performs better than other methods consistently.

\subsection{Offline Experimental Results on Different Continual Learning Methods}
Following the common practice in CTR prediction research, we use log-loss and AUC score \cite{fawcett2006introduction} to measure a model's prediction ability and ranking quality. The lower the log-loss is and the higher and AUC score is, the better the model is. Due to the large scale of our data, all experiments are conducted with TensorFlow on a distributed computing platform and are run under the same hyperparameters such as batch size and learning rate.

We report the continuous performance of the last four days' data. As Table \ref{table:AUC-logloss} shows, from time $T$ to time $T+3$, in terms of AUC score, COLF always beats baseline model by a large margin. The relative improvement of AUC score is 1.01\%, 0.99\%, 0.97\%, 1.03\% respectively. Note that 0.1 percent improvement of AUC is significant enough in practice for business growth. Log-loss of COLF also is lower than baseline in all four days. This result shows that COLF gives more accurate CTR prediction and has higher ranking quality than baseline model, and the improvement is robust over time.

Table \ref{table:AUC-logloss} also shows that COLF outperforms ADER and CBRS consistently. The average relative AUC gains are 0.90\% and 0.47\% respectively. In contrast to ADER which fills its memory by weighted sampling strategy whose weights are determined by historical frequency, COLF populates its memory by explicitly controlling the gap between memory and target data and thus enables stronger positive knowledge transfer from the past to the future. The relative weak performance of CBRS is expected since its memory population strategy aims at solving class imbalance problem with no emphasis on the input data's distribution.
\begin{table*}[ht]
\caption{Comparison on different continual learning techniques using real world data. BaseModel is a vanilla DSIN model with no continual learning technique. COLF outperforms all other competitors in all time steps from $T$ to $T+3$.}
\label{table:AUC-logloss}
\begin{center}
\begin{sc}
\begin{tabular}{lccccccccccc}
\toprule
& \multicolumn{2}{c}{$t=T$} & & \multicolumn{2}{c}{$t=T+1$} & & \multicolumn{2}{c}{$t=T+2$} & & \multicolumn{2}{c}{$t=T+3$} \\
\cline{2-3}  \cline{5-6} \cline{8-9}  \cline{11-12} 
Model & logloss & AUC & & logloss & AUC & & logloss & AUC & & logloss & AUC  \\
\toprule
BaseModel   &0.1399 &  0.7450 & & 0.1411& 0.7418 && 0.1407 &0.7452&& 0.1395 & 0.7483 \\
CBRS & 0.1399 & 0.7458 & &  0.1405 & 0.7430 &&0.1401 & 0.7458&&0.1395 & 0.7484\\
ADER  & 0.1395 & 0.7489 & &  0.1404 & 0.7455 &&0.1398 & 0.7490 &&0.1393&0.7514 \\
COLF   & 0.1388 & 0.7525 &  &  0.1395&  0.7491 && 0.1391 & 0.7524 && 0.1386 & 0.7560\\ 
\bottomrule
\end{tabular}
\end{sc}
\end{center}
\end{table*}

\subsection{Offline Experimental Results On Different CTR Models with COLF}
We investigate the performance of COLF with different CTR base models in this section. The three selected base CTR prediction models are as follows.

\textbf{Logistic Regression.} Logistic regression(LR) was  widely used for CTR prediction task before the thriving of deep neural models. It could be seen as a shallow neural network with only one dense layer. It is the weakest baseline here.

\textbf{Wide\&Deep}. Wide\&Deep is an embedding based neural model for CTR prediction and has achieved superior performance over traditional non-neural models \cite{cheng2016wide}. We include it here as the neural model's baseline.

\textbf{DSIN.} DSIN is the state-of-the-art attention-based neural model \cite{feng2019deep}. It takes advantages of user behavior data and sequence modeling to achieve better modeling of user interest. We include it here as the strongest baseline.

As Table \ref{table:differentCTRModels} shows, CTR base models with COLF always have better performance than those without COLF. Note that AUC gain of DSIN+COLF over DSIN is larger than that of   Wide\&Deep+COLF over Wide\&Deep or that of LR+COLF over LR. We conjecture that with more complex feature space (e.g., DSIN with user behavior features), the data has a subtler pattern of distribution drifting, thus COLF is able to contribute more.
\begin{table}[ht]
\caption{Performance comparison on different CTR models with or without COLF. All the lines with COLF calculate the relative AUC gain by comparing with the non-COLF competitor.}
\label{table:differentCTRModels}
\begin{center}
\begin{small}
\begin{sc}
\begin{tabular}{lccr}
\toprule
Model & log-loss & AUC & Gain \\
\midrule
LR    & 0.1447 & 0.7043 &  - \\
LR+COLF & 0.1444  & 0.7084 & 0.58\%  \\
Wide\&Deep& 0.1431 & 0.7150 & - \\
Wide\&Deep+COLF & 0.1426 & 0.7196 & 0.64\% \\
DSIN   & 0.1399 & 0.7450 &  - \\
DSIN+COLF &0.1388 & 0.7525 & 1.10\% \\
\bottomrule
\end{tabular}
\end{sc}
\end{small}
\end{center}
\end{table}
\subsection{Ablation Study}
Table \ref{table:ablationstudy} shows that removing any part of COLF leads to worse performance. It is because that removing either relevant memory or new memory causes  useful knowledge forgetting or insufficient knowledge transfer, while keeping all old memory suffers from negative knowledge transfer. Note that COLF without modular architecture performs much worse. We conjecture it is because that a portion of data in the memory will be consumed by $\hat{f}_t$ multiple times  in this setting, causing several over-fitting.
\begin{table}[ht]
\caption{Performance comparison on different variants of COLF.}
\label{table:ablationstudy}
\begin{center}
\begin{small}
\begin{sc}
\begin{tabular}{lr}
\toprule
Model  & AUC \\
\midrule
DSIN with COLF & 0.7525  \\
w/o modular architecture  & 0.7479\\
w/o old memory discarding & 0.7501\\
w/o relevant memory  & 0.7510 \\
w/o new memory  & 0.7510\\
\bottomrule
\end{tabular}
\end{sc}
\end{small}
\end{center}
\end{table}

\subsection{Online Experimental Results}
\begin{figure}[ht]
    \begin{center}
    \includegraphics[width=\linewidth]{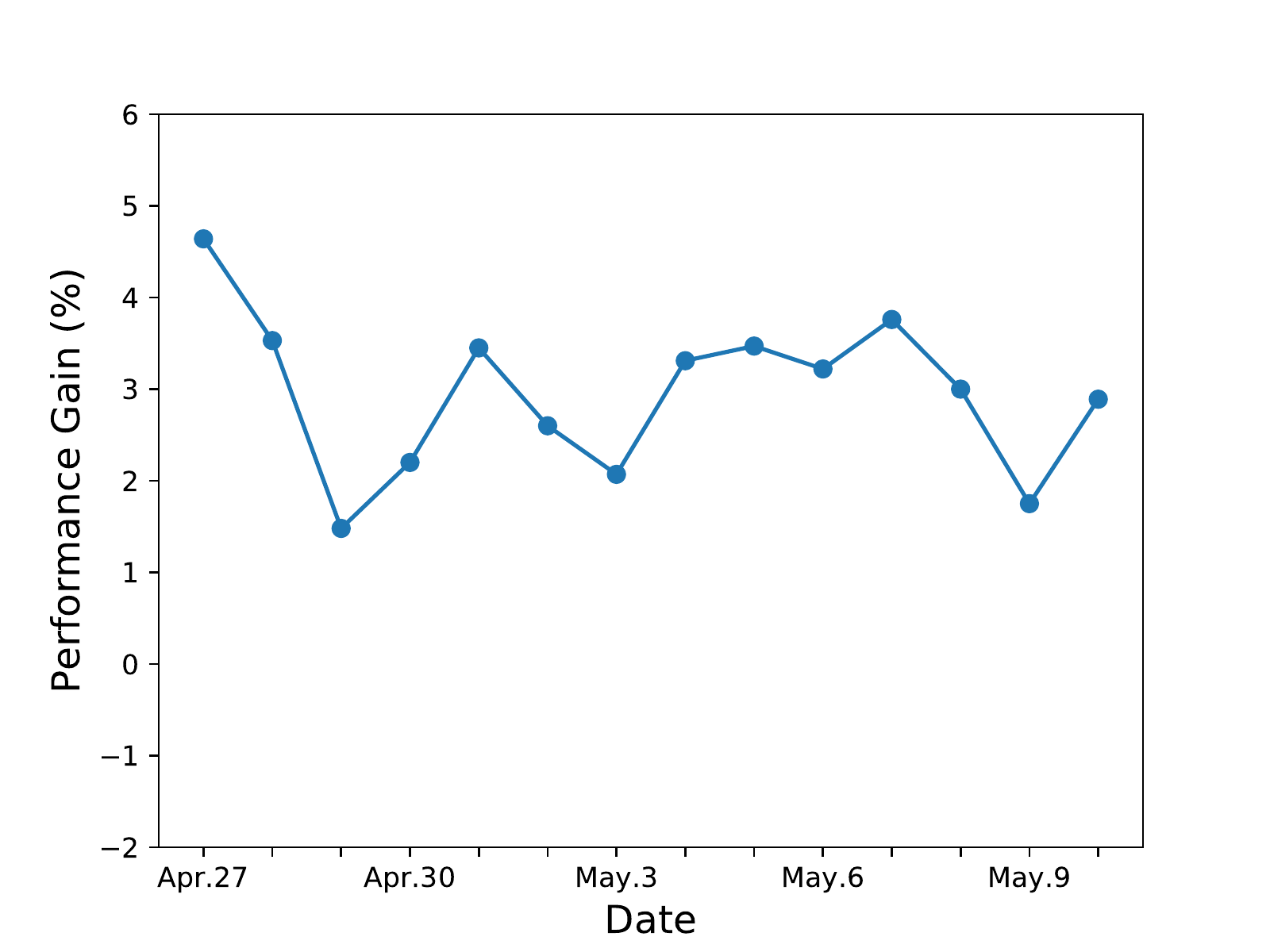}
    \caption{Online COLF model's CTR gain over baseline between 2020 April 27th and 2020 May 10th. The gain is always positive and is statistically significant at the 0.05 level.}
    \label{fig:onlineExpMayPerformanceGain}
    \end{center}
\end{figure}

We have deployed COLF in production  and now it is serving the main traffic of a major mobile shopping  app's advertising system in China. The online baseline is a highly optimized DSIN model. Carefully designed online A/B test on the advertising system was conducted. During the whole test period, COLF contributed up to 2.89\% CTR promotion and 2.52\% RPM(Revenue per thousand impressions, an index to measure the efficiency of traffic monetization) boost. Note that the advertising system serves millions of users every day and just 1\% gain  leads to significant revenue growth. As is illustrated in Figure \ref{fig:onlineExpMayPerformanceGain}, the daily boost is quite robust even during holidays (May 1st is International Workers' Day when there is a seven-day long holiday in China).

\section{Conclusion}
\label{section:conclusion}
We studied continual learning for CTR prediction to address the non-stationary and drifting problem of click data in this work. We gave a formal formulation of the problem and proposed COLF, a hybrid approach that marries memory replay with a modular architecture to foster positive knowledge transfer and mitigate catastrophic forgetting. Both offline and online experiments demonstrated COLF's superiority over existing methods. An interesting extension of our work is the continual learning problem with delayed feedback, which is useful for conversion rate prediction in advertising systems. We plan to investigate it in the future.

\bibliographystyle{IEEEtran}
\bibliography{ref}
\end{document}